\documentstyle[preprint,eqsecnum,aps]{revtex}
\begin{document}
\def\appls{\hbox{$<$\kern-.75em\lower 1.00ex\hbox{$\sim$}}}
\draft
\title{STUDY OF RESONANT STRUCTURE OF $S$-WAVE\\
AMPLITUDES IN $\pi^- p_\uparrow \to \pi^- \pi^+ n$\\
MEASURED ON POLARIZED TARGET AT 17.2 GeV/c\\
IN THE MASS RANGE OF 580--1080 MeV.}
\author{M. Svec\footnote{electronic address: svec@hep.physics.mcgill.ca}}
\address{Physics Department, Dawson College, Montreal, Quebec,  
Canada H3Z 1A4\\
and\\
McGill University, Montreal, Quebec, Canada H3A 2T8}
\maketitle
\begin{abstract}
We have performed a new model independent amplitude analysis of  
reaction $\pi^- p \to \pi^- \pi^+ n$ measured at CERN at 17.2 GeV/c  
on polarized target. The analysis extends the range of dipion mass  
from 600--880 MeV used in the previous study to 580--1080 MeV. Our  
purpose is to study the role of $f_0(980)$ interference and to  
better determine the resonance parameters of $\sigma(750)$ resonance  
from fits to both $S$-wave amplitudes $|\overline S|^2\Sigma$ and  
$|S|^2\Sigma$ with recoil nucleon transversities ``up'' and  
``down'', respectively. This work is the first attempt to determine  
resonance parameters of a resonance from Breit-Wigner fits to  
production amplitudes of opposite transversity. We used several  
fitting procedures to understand the resonant structure of the  
$S$-wave amplitudes in this mass range. First we performed separate  
fits to $|\overline S|^2\Sigma$ and $|S|^2\Sigma$ using a  
Breit-Wigner parametrization for $\sigma(750)$ and $f_0(980)$, and a  
constant coherent background. We obtained an excellent fit to the  
data with a low $\chi^2$ per data point $(\chi^2/dpt)$. However the  
width of $\sigma$ is narrow 110--76 MeV in $|\overline S|^2\Sigma$  
while it is broad 217--233 MeV in $|S|^2\Sigma$. To decide whether  
the apparent dependence of $\Gamma_\sigma$ on nucleon spin is a real  
physical effect or an artifact of separate fits, we performed  
simultaneous fits to $|\overline S|^2\Sigma$ and $|S|^2\Sigma$.  
First we assumed a common single $\sigma$ pole. We found a good fit  
with $m_\sigma=775\pm 17$ MeV, $\Gamma_\sigma = 147 \pm 33$ MeV but  
the $\chi^2/dpt$ is larger than $\chi^2/dpt$ from separate fits. We  
concluded that the results of separate fits may indicate existence  
of two $\sigma$ poles, one with a narrow and the other with a broad  
width. Self-consistency requires that both poles contribute to both  
$S$-wave amplitudes. The simultaneous fit to $|\overline S|^2\Sigma$  
and $|S|^2\Sigma$ with two common $\sigma$ poles yields an  
excellent fit with a lower $\chi^2/dpt$ than the separate fits. All  
amplitudes are dominated by a $\sigma$ state with $m_\sigma = 786  
\pm 24$ MeV and $\Gamma_\sigma = 130\pm 47$ MeV. All amplitudes  
receive a weaker contribution from a narrower $\sigma^\prime$ state  
with $m_{\sigma^\prime} = 670 \pm 30$ MeV and  
$\Gamma_{\sigma^\prime} = 59 \pm 58$ MeV. We propose to identify  
$\sigma^\prime(670)$ and $\sigma(786)$ with colour-electric and  
colour-magnetic modes of lowest mass scalar gluonium $0^{++} (gg)$.

\end{abstract}
\section{Introduction}
In 1978, Lutz and Rybicki showed\cite{lutz78} that almost complete  
amplitude analysis of reactions $\pi N \to \pi^+ \pi^- N$ and $KN  
\to K^+ \pi^- N$ is possible from measurements in a single  
experiment on a transversely polarized target. More recently it was  
shown\cite{svec97,svec96} that amplitude analyses of reactions  
$\pi^- p \to \pi^0 \pi^0 n$ and $\pi^- p \to \eta \pi^- p$ are also  
possible from measurements on transversely polarized target. The  
work of Lutz and Rybicki opened a whole new approach to hadron  
spectroscopy by enabling to study the production of resonances on  
the level of spin amplitudes rather than spin-averaged  
cross-sections.

In amplitude analyses of these reactions resonances are observed in  
mass distribution of recoil nucleon transversity amplitudes  
$|\overline A |^2\Sigma$ and $|A|^2\Sigma$ corresponding to recoil  
nucleon spin ``up'' or ``down'' relative to the scattering plane. In  
traditional hadron spectroscopy, the nucleon spin is irrelevant and  
resonance should appear in both spin amplitudes with the same mass,  
with the same width and with similar mass distributions. The  
measurements of reactions $\pi^- p \to \pi^- \pi^+ n$, $\pi^+ n \to  
\pi^+ \pi^- p$ and $K^+ n \to K^+ \pi^- P$ at CERN on polarized  
targets seem to challenge these expectations. These CERN experiments  
reveal shapes of resonant mass distributions in amplitudes of  
opposite transversity that are different and dependent significantly  
on the nucleon  
spin\cite{groot78,becker79,becker79b,chabaud83,sakrejda84,rybicki85,kamins97,lesquen85,lesquen89,svec90,svec92,svec92b,svec92c,svec96b,svec97b}.

An important advantage of amplitude analyses is that new resonances  
can be detected on the level of spin amplitude that are not visible  
in the spin-averaged measurements on unpolarized targets. Such is  
the case of the new scalar resonance $\sigma(750)$ found in  
amplitude analyses of reactions $\pi^- p\to\pi^- \pi^+ n$ at 17.2  
GeV/c at low momentum transfers $-t =$ 0.005--0.2 (GeV/c)$^2$ and in  
$\pi^+ n \to \pi^+ \pi^- p$ at 5.98 and 11.85 GeV/c at larger  
momentum transfers $-t=$ 0.2--0.4  
(GeV/c)$^2$\cite{svec92c,svec96b,svec97b}. Using Monte Carlo method  
of amplitude analysis, it was found in \cite{svec96b} that in $\pi^-  
p\to \pi^- \pi^+ n$ reaction both solutions for spin ``up''  
$S$-wave amplitude $|\overline S|^2\Sigma$ resonate below 880 MeV  
while both solutions for spin ``down'' $S$-wave amplitude  
$|S|^2\Sigma$ appeared nonresonating. This result is in perfect  
agreement with $\chi^2$ minimization method of amplitude analysis  
used in \cite{groot78,becker79} (see Fig.~1 and 2 of  
\cite{svec97b}). Detailed quantitative Breit-Wigner fits to  
$|\overline S|^2\Sigma$ in \cite{svec97b} showed the importance of  
coherent background for determination of the resonance parameters of  
$\sigma(750)$. The best fit in \cite{svec97b} gives $m_\sigma=753  
\pm 19$ MeV and $\Gamma_\sigma = 108 \pm 53$ MeV. The excellent  
agreement of these values for $m_\sigma$ and $\Gamma_\sigma$ with  
the Ellis-Lanik theorem\cite{ellis85} strongly suggests that  
$\sigma(750)$ is the lowest mass scalar gluonium\cite{svec97b}.

T\"ornqvist expressed concern\cite{tornquist,partphys96} that the  
low mass and the narrow width of $\sigma(750)$ is due to the neglect  
of interference of $\sigma(750)$ with $f_0$ (980) in our  
Breit-Wigner fits which initially extended only to 880 MeV in dipion  
mass using the data set of Ref.~[5]. The apparent nonresonating  
behaviour of the amplitude $|S|^2\Sigma$ below 880 MeV was also  
puzzling. To investigate these questions we have performed a new  
amplitude analysis of $\pi^- p \to \pi^- \pi^- n$ reaction at 17.2  
GeV/c and small momentum transfers $-t=$0.005--0.20 (GeV/c)$^2$  
extending the dipion mass range up to 1080 MeV using the data set of  
Ref.~[7].

In this report we present the results of our detailed study of  
resonant structure of $S$-wave amplitudes $|\overline S|^2\Sigma$  
and $|S|^2\Sigma$ in the mass region of 580--1080 MeV. This work is  
the first attempt to determine resonance parameters of a resonance  
from fits to production amplitudes of opposite nucleon transversity.  
All previous determinations of resonance parameters used  
spin-averaged mass distributions or, in the case of Ref.~[18], only  
one spin amplitude $(|\overline S|^2\Sigma)$.

The new results for $|\overline S|^2\Sigma$ shows a narrow  
resonance below 880 MeV, followed by an enhancement above 900 MeV  
and a dramatic dip at $\sim$1000 MeV corresponding to $f_0(980)$.  
The amplitude $|S|^2\Sigma$ shows a broader structure with a dip at  
$\sim$1000 MeV.

First we performed separate fits to $|\overline S|^2\Sigma$ and  
$|S|^2$ using a coherent sum of $\sigma(750)$ and $f_0(980)$  
Breit-Wigner amplitudes and constant background. We find that  
$\sigma$ contributes to both amplitudes but with different widths.  
Solutions 1 and 2 of amplitude $|\overline S|^2\Sigma$ have a narrow  
width of 100 and 76 MeV, respectively. Solutions 1 and 2 of  
amplitude $|S|^2\Sigma$ have a broader width of 217 and 233 MeV. The  
interference with $f_0(980)$ is crucial in reproducing the  
enhancements above 900 MeV and the dip at 1010 MeV.

The narrow width in the spin ``up'' amplitude $|\overline  
S|^2\Sigma$ and the broad width in the spin ``down'' amplitude  
$|S|^2\Sigma$ is inconsistent with the traditional view of a  
resonace as a particle with single mass and width. To see if we can  
describe the mass spectra of $|\overline S|^2\Sigma$ and  
$|S|^2\Sigma$ with a single $\sigma$ with the same resonance  
parameters, we have performed a simultaneous fit to $|\overline  
S|^2\Sigma$ and $|S|^2\Sigma$ hoping that the differences in the  
resonant shapes can be accounted for by differences in the  
background. We obtained an acceptable fit but the $\chi^2$ per data  
point $(\chi^2/dpt)$ was significantly higher than $\chi^2/dpt$ in  
separate fits. We thus obtained a poorer fit.

Our next step was to admit that a narrow $\sigma$ contributes to  
$|\overline S|^2\Sigma$ and a broad $\sigma$ contributes to  
$|S|^2\Sigma$. But a self-consistency requires that the narrow  
$\sigma$ contributes also to $|S|^2\Sigma$, and the broad $\sigma$  
contributes also to $|\overline S|^2\Sigma$. We expected these  
additional contributions to be small in a simultaneous two-pole fit  
to $|\overline S|^2\Sigma$ and $|S|^2\Sigma$. The results of the fit  
were surprising. Our best fit had $\chi^2/dpt$ better than the  
separate fits to $|\overline S|^2\Sigma$ and $|S|^2\Sigma$. And the  
fit required in all amplitudes the presence of two $\sigma$  
resonances: higher mass and broader $\sigma$ with average mass  
$m_\sigma = 786 \pm 24$ MeV and $\Gamma_\sigma = 130 \pm 47$, and a  
lower mass and narrower $\sigma^\prime$ with average mass of  
$m_{\sigma^\prime} = 670 \pm 30$ MeV and $\Gamma_{\sigma^\prime} =  
59 \pm 58$ MeV. In all $S$-wave amplitudes the higher mass $\sigma$  
dominates the $\sigma^\prime$ contribution. Since this solution  
gives a very low $\chi^2/dpt$, it must be taken seriously. We  
propose to identify the $\sigma^\prime(670)$ and $\sigma(786)$ as  
color electric and color magnetic modes of lowest mass scalar  
gluonium $0^{++} (gg)$, respectively, following the gluonium  
classification scheme of Bjorken\cite{bjorken79,mosher80,bjorken80}  
elaborated in\cite{flamm82}.

The paper is organized as follows. In Section II we present and  
discuss the results of our new amplitude analysis. In Section III we  
introduce the parametrization of unnormalized $S$-wave amplitudes  
with a single $\sigma$ pole and present the results of separate fits  
to amplitudes $|\overline S|^2\Sigma$ and $|S|^2\Sigma$.  
Simultaneous fits to these $S$-wave amplitudes using a single  
$\sigma$ pole are presented in Section IV. In Section V we introduce  
the parametrization with two $\sigma$ poles and discuss the results  
of simultaneous fits to $|\overline S|^2\Sigma$ and $|S|^2\Sigma$  
using this parametrization. In Section VI we discuss the possible  
constituent nature of the two $\sigma$ states $\sigma^\prime(670)$  
and $\sigma(786)$ resulting from the best fit. The paper closes with  
a summary in Section VII.

\section{Amplitude analysis}

The high statistics CERN-Munich measurement of $\pi^-  
p\to\pi^-\pi^+ n$ at 17.2 GeV/c on polarized target was reported in  
four data sets with kinematics given in the following Table:

\narrowtext
\begin{quasitable}
\begin{tabular}{lcccc}
Set & $-t$ & $m$ & $\Delta m$ & Ref.\\
&(GeV/c)$^2$ & (MeV) & (MeV)\\
\tableline
1 & 0.005--0.20 & 600--900 & 20 & [4,5]\\
2 & 0.01--0.20 & 580--1780 & 40 & [6]\\
3 & 0.005--0.20 & 580--1600 & 20 & [7,10]\\
4 & 0.20--1.00 & 620--1500 & 40 & [8,9]\\
\end{tabular}
\end{quasitable}

\noindent
In the Table $\Delta m$ is the size of the mass bins. There is only  
one $t$-bin covering the whole interval of indicated $-t$.  
Only\cite{groot78,becker79} report the results for amplitudes for  
the Set 1. In\cite{becker79b} and\cite{kamins97} ratios  
$|S|/|\overline S|$ and $S$-wave intensity $I_S$ are given, so it is  
possible to reconstruct approximately amplitudes $|S|^2\Sigma$ and  
$|\overline S|^2\Sigma$ for the Sets 2 and 3. These are shown in  
Fig.~5 of Ref.~[18] and Fig.~2 below, respectively. All amplitude  
analysis\cite{groot78,becker79,becker79b,chabaud83,sakrejda84,rybicki85,kamins97}  
use $\chi^2$ minimization method to find solution for amplitudes  
and their errors.

For invariant masses below 1000 MeV, the dipion system in reactions  
$\pi N \to \pi^+\pi^- N$ is produced predominantly in spin states  
$J=0$ ($S$-wave) and $J=1$ ($P$-wave). The experiments on polarized  
targets then yield 15 spin-density-matrix (SDM) elements, or  
equivalently 15 moments, describing the dipion angular  
distribution\cite{lutz78,svec92}. The measured normalized  
observables are expressed in terms of two $S$-wave and six $P$-wave  
normalized nucleon transversity amplitudes. In our normalization

\begin{equation}
|S|^2 + |\overline S|^2 + |L|^2 + |\overline L|^2 + |U|^2 +  
|\overline U|^2 + |N|^2 + |\overline N|^2 = 1
\end{equation}

\noindent
where $A=S,L,U.N$ and $\overline A = \overline S, \overline L,  
\overline U, \overline N$ are the normalized nucleon transversity  
amplitudes with recoil nucleon transversity ``down'' and ``up''  
relative to the scattering plane. The $S$-wave amplitudes are $S$  
and $\overline S$. The $P$-wave amplitudes $L$, $\overline L$ have  
dimeson helicity $\lambda=0$ while the pairs $U$, $\overline U$ and  
$N$, $\overline N$ are combinations of amplitudes with helicities  
$\lambda = \pm 1$ and have opposite $t$-channel-exchange  
naturality\cite{lutz78,svec92}. The unnatural exchange amplitudes  
receive contributions from ``$\pi$'' and ``$A_1$'' exchanges. The  
natural exchange amplitudes $N$, $\overline N$ are both dominated by  
``$A_2$'' exchange.

Amplitude analysis expresses analytically the eight normalized moduli

\begin{equation}
|S|, |\overline S|, |L|, |\overline L|, |U|, |\overline U|, |N|,  
|\overline N|
\end{equation}

\noindent
and six independent cosines of relative phases

\begin{equation}
\cos (\gamma_{SL}), \cos (\gamma_{SU}), \cos(\gamma_{LU})
\end{equation}

\[
\cos (\overline\gamma_{SL}), \cos (\overline\gamma_{SU}), \cos  
(\overline\gamma_{LU})
\]

\noindent
in terms of the measured observables\cite{lutz78,svec92}. In (2.3)  
$\cos\gamma$ and $\cos\overline\gamma$ are cosines of relative  
phases between pairs of amplitudes with opposite transversity (e.g.  
between $S$ and $L$, and between $\overline S$ and $\overline L$).

There are two similar solutions in each $(m,t)$  
bin\cite{lutz78,svec92}. However in many $(m,t)$ bins the solutions  
are unphysical: either a cosine has magnitude larger than 1 or the  
two solutions for moduli are complex conjugate with a small  
imaginary part. Unphysical solutions also complicate error analysis.  
Two methods are used to find physical solutions for amplitudes and  
their errors. They are $\chi^2$ minimization method and Monte Carlo  
method. The two methods are described  
in\cite{groot78,svec96b,svec97b}.

In Ref.~[17] we used Monte Carlo method for finding the physical  
solutions for amplitudes and their errors in $\pi^- p \to \pi^-  
\pi^+ n$ using the data Set 1 for dipion masses in the range  
600--900 MeV. In Ref.~[18] we show that the results obtained using  
Monte Carlo method agree with the results using $\chi^2$  
minimization method but are smoother and give consistently lower  
$\chi^2/dpt$ in Breit-Wigner fits.

In this work we report results of Monte Carlo amplitude analysis of  
$\pi^- p \to \pi^- \pi^+ n$ reaction using the data Set 3 in the  
mass range 580--1080 MeV in order to study the  
$\sigma(750)-f_0(980)$ interference. Above 1000 MeV our analysis is  
only approximate as we neglect the $D$-wave. However, below 1080 MeV  
the $D$-wave contribution is still small and its neglect does not  
affect the structure of $S$-wave amplitudes. This is confirmed by  
direct comparison with $S$-wave amplitudes obtained  
in\cite{kamins97} using the $\chi^2$ method with $D$-wave included  
above 980 MeV (see Fig.~2d below). The Monte Carlo method is based  
on 40,000 selections of spin-density matrix elements within their  
errors in each $(m,t)$ bin. No physical solution was found in 3 bins  
-- 650 MeV, 850 MeV and 930 MeV.

The amplitude analysis is carried out for normalized amplitudes  
$|A|^2$ and $|\overline A|^2$, $A = S,L,U,N$. However, from the  
spectroscopic point of view the relevant information is contained in  
the unnormalized amplitudes $|A|^2\Sigma$ and $|\overline  
A|^2\Sigma$ where $\Sigma \equiv d^2\sigma/dmdt$ is the reaction  
cross-section. It is the unnormalized amplitudes $|A|^2\Sigma$ and  
$|\overline A|^2\Sigma$ which represent the direct spin-dependent  
contributions to the mass distribution $d^2\sigma/dmdt$. The  
unnormalized moduli $|A|^2\Sigma$ and $|\overline A|^2\Sigma$ were  
calculated using the $\Sigma = d^2\sigma/dmdt$ from Fig.~12 of  
Ref.~[26].

The results of our new Monte Carlo amplitude analysis of $\pi^- p  
\to \pi^-\pi^+ n$ in the mass range 580--1080 MeV are shown in  
Fig.~1. The first thing we observe is that the $\rho^0$ peak in  
$d\sigma/dmdt$ is not uniformly reproduced in all $P$-wave  
amplitudes. In fact, the amplitudes $|\overline N|^2\Sigma$ and  
$|U|^2\Sigma$ show considerable suppression of $\rho^0$ production.  
The shapes of mass distributions with opposite nucleon  
transversities show considerable differences in both $S$- and  
$P$-wave amplitudes. The amplitudes $|S|^2\Sigma$, $|L|^2\Sigma$,  
$|U|^2\Sigma$ with recoil nucleon transversity ``down'' are smaller  
and broader than the amplitudes $|\overline S|^2\Sigma$, $|\overline  
L|^2\Sigma$ and $|\overline U|^2\Sigma$ with recoil nucleon  
transversity ``up''. The opposite is true for the natural exchange  
amplitudes $|\overline N|^2\Sigma$ and $|N|^2\Sigma$.

The $S$-wave amplitude $|\overline S|^2\Sigma$ shows a resonant  
behaviour below 900 MeV in both solutions. The relative phase  
$\overline\gamma_{SL}$ between $\overline S$ and $\overline L$  
amplitudes is near zero in Solution 1 and a small constant in  
Solution 2. Since $|\overline L|^2\Sigma$ clearly resonates, the  
amplitude $|\overline S|^2\Sigma$ must resonate as well. The  
amplitudes $|S|^2\Sigma$ and $|L|^2\Sigma$  are both broader and  
their relative phase $\gamma_{SL}$ is again near zero in Solution 2.  
This strongly suggests that the amplitude $|S|^2\Sigma$ also  
resonates.

Both solutions in both $S$-wave amplitudes dip at 1010 MeV. This  
dip is accompanied with a dramatic change of phase  
$\overline\gamma_{SL}$ and $\overline\gamma_{SU}$ above 1000 MeV.  
This behaviour is interpreted as evidence for a narrow resonance  
$f_0(980)$. Its contribution must be taken into account in fits to  
amplitudes $|\overline S|^2\Sigma$ and $|S|^2\Sigma$.

In Figure 2 we summarize the results for the $S$-wave amplitudes.  
Fig.~2a and 2b are based on Monte Carlo method of amplitude analysis  
of data Set 1 and 3, respectively. The amplitude $|\overline  
S|^2\Sigma$ clearly resonates below 880 MeV in both solutions but  
shows an enhancement between 880 MeV and 980 MeV which is  
particularly large on Solution 2. The enhancement is followed by a  
rapid drop at 990 MeV with a dip at 1010 MeV. The apparent broad  
structure of $|S|^2\Sigma$ amplitude is due to poor separation of  
the resonant structure below 880 MeV and the enhancement between 880  
and 980 MeV which are of comparable size. The rapid drop of the  
enhancement at 990 MeV and dip at 1010 MeV are clearly seen even in  
$|S|^2\Sigma$ amplitude. As we shall see later the enhancement is  
due to interference of $f_0(980)$ with coherent background.

Fig.~2c and 2d are based on $\chi^2$ minimization methods of  
amplitude analysis of data Sets 1 and 3, respectively. Fig.~2c is  
based on Fig.~10 of Ref.~[5] and Fig.~2d is based on Fig.~2 of  
Ref.~[10]. The results are consistent with the Monte Carlo amplitude  
analysis.

The spin-averaged $S$-wave intensity is defined as $I_S = (|S|^2 +  
|\overline S|^2)\Sigma$. Since there are two independent solutions  
for the amplitudes $|S(i)|^2$ and $|\overline S(j)|^2$, $i,j=1,2$,  
we obtain 4 solutions for $I_S$ which we label as follows

\begin{equation}
I_S (i,j) = (|S(i)|^2 + |\overline S(j)|^2)\Sigma
\end{equation}

\noindent
The results are shown in Fig.~3. We see that only solutions  
$I_S(1,1)$ and $I_S(2,1)$ are clearly resonating while the solutions  
$I_S(1,2)$ and $I_S(2,2)$ do not have a clear resonant structure  
due to enhancement in the range 880--980 MeV. It is also for this  
reason that it is better to study directly the resonant structure of  
the amplitudes $|\overline S|^2\Sigma$ and $|S|^2\Sigma$ rather  
than their spin averaged $S$-wave intensity.

\section{Separate fits to amplitudes $|\overline S|^2\Sigma$ and  
$|S|^2\Sigma$}

\subsection{Parametrization of amplitudes}

To understand the resonant structure of the $S$-wave amplitudes  
$|\overline S|^2\Sigma$ and $|S|^2\Sigma$ we first performed  
separate fits to these amplitudes using a parametrization which  
included single $\sigma(750)$ Breit-Wigner pole, $f_0(980)$  
Breit-Wigner pole and a coherent background. Following Ref.~[18], we  
write the unnormalized $S$-wave amplitudes in the form

\begin{equation}
|\overline S|^2\Sigma = q |\overline F|^2
\end{equation}

\[
|S|^2\Sigma = q |F|^2
\]

\noindent
where $q$ is the mass dependent phase space factor determined in  
[18] to be simply the c.m.s. momentum of $\pi^-$ in the $\pi^-\pi^+$  
rest frame. The amplitudes $\overline F$ and $F$ are unnormalized  
amplitudes which are parametrized in terms of resonance  
contributions and coherent background. For amplitude $F$ we write

\begin{equation}
F = R_\sigma (s,t,m) BW_\sigma(m) + R_f (s,t,m) BW_f (m) + Q (s,t,m)
\end{equation}

\noindent
where $BW_R$ is the Breit-Wigner amplitude

\begin{equation}
BW_R = {{m_R \Gamma}\over{m_R^2 - m^2 - im_R \Gamma}}
\end{equation}

\noindent
where $m_R$ is the resonant mass, $R = \sigma, f$. In the following  
$f$ will refer always to $f_0(980)$ resonance. The mass dependent  
width $\Gamma(m)$ depends on spin $J$ and has a general form

\begin{equation}
\Gamma = \Gamma_R ({q\over q_R})^{2J+1} {{D_J (q_R r)}\over{D_J (qr)}}
\end{equation}

\noindent
In (3.4) $q_R = q (m = m_R)$ and $D_J$ are the centrifugal barrier  
functions of Blatt and Weishopf\cite{blatt52}

\begin{equation}
D_0 (qr) = 1.0
\end{equation}

\[
D_1 (qr) = 1.0 + (qr)^2
\]

\noindent
where $r$ is the interaction radius. We recall that

\[
Re\  BW_R = ({{m^2_R - m^2}\over{m_R\Gamma}}) |BW_R|^2 \equiv w_R  
|BW_R|^2\
\]

\begin{equation}
Im\ BW_R = |BW_R|^2
\end{equation}

\noindent
In (3.2) the term $Q(s,t,m)$ is the coherent nonresonant  
background. The amplitude $\overline F$ has a form similar to (3.2)  
with obvious replacements $R_\sigma \to \overline R_\sigma$, $R_f  
\to \overline R_f$ and $Q\to \overline Q$.

The energy variable $s$ is fixed and will be omitted in the  
following. Since the experimental mass distributions are averaged  
over a broad $t$-bin $-\Delta t = 0.005$--0.20 (GeV/c)$^2$, we will  
work with amplitudes averaged over the same $t$ interval. The  
$t$-averaged amplitudes have the same form as (3.2), so we will  
simply assume that $R_\sigma$, $R_f$ and $Q$ are all  
$t$-independent. We will also assume that $R_\sigma$, $R_f$ and $Q$  
depend only weakly on the dipion mass $m$ and can be taken as  
constants in our fits. It is then convenient to factor out the phase  
of $R_\sigma$ and define

\begin{equation}
R_\sigma = |R_\sigma|e^{i\phi} = \sqrt N_S e^{i\phi}
\end{equation}

\[
C=R_f/R_\sigma = C_1 + iC_2
\]

\[
B = Q/R_\sigma = B_1 + iB_2
\]

\noindent
Then the unnormalized amplitude (3.2) has the form

\begin{equation}
F = \sqrt N_S \{ BW_\sigma + CBW_f + B\} e^{i\phi}
\end{equation}

\noindent
and the parametrization of $|S|^2\Sigma$ reads

\[
|S|^2\Sigma = q N_S\{ [1 + 2w_\sigma B_1 + 2B_2] |BW_\sigma|^2 +
\]

\[
+ B_1^2 + B_2^2 + [C_1^2 + C_2^2] |BW_f|^2 +
\]

\[
+ 2 [(w_\sigma |BW_\sigma|^2 + B_1) (w_f C_1 - C_2 ) +
\]

\begin{equation}
+ (|BW_\sigma|^2 + B_2) (C_1 + w_f C_2)] |BW_f|^2\}
\end{equation}

\noindent
The parametrization of $|\overline S|^2\Sigma$ has the same form as  
(3.9).

\subsection{Results of separate fits to $|\overline S|^2\Sigma$ and  
$|S|^2\Sigma$}

The measured mass distributions $|\overline S|^2\Sigma$ and  
$|S|^2\Sigma$ were augmented at 650 and 850 MeV mass bins by data  
points from Monte Carlo amplitude analysis using the data Set  
1\cite{svec96b}. Independent fits were performed to both solutions  
of both $S$-wave amplitudes with mass and width of $\sigma$  
resonance being free parameters. The fitting was done using the CERN  
optimization program FUMILI\cite{silin85}. Initially the $f_0  
(980)$ was fixed with mass $m_f = 980$ MeV and width $\Gamma_f = 48$  
MeV. An improved fit was obtained when $m_f$ and $\Gamma_f$ were  
allowed to be free parameters to be fitted. The best fit was  
obtained with fitted $f_0(980)$ using as initial values of  
parameters $B_i, C_i, i=1,2$ the output values from the fit with  
fixed $f_0 (980)$. The results are shown in Fig.~4a. The fitted  
values of $m_\sigma, \Gamma_\sigma, m_f$ and $\Gamma_f$ and the  
associated $\chi^2/dpt$ ($\chi^2$ per data point) for each solution  
are given in Table I.

We see from Fig.~4a that the parametrization (3.9) reproduces well  
the structure of $S$-wave amplitudes $|\overline S|^2\Sigma$ and  
$|S|^2\Sigma$ in both solutions. The $\sigma$ resonance peaks are  
clearly visible in both solutions to $|\overline S|^2\Sigma$ while  
the presence of $\sigma$ resonance in $|S|^2\Sigma$ is required to  
explain the broad structure observed in this amplitude. The  
interference of $f_0(980)$ with coherent background and the $\sigma$  
resonance is responsible for the enhancements in the mass region  
880--980 MeV, the rapid decrease at 990 MeV and the dip at 1010 MeV  
in all solutions.

We see from the Table I that the masses $m_\sigma$ are similar in  
all solutions. We might expect the same situation for the width  
$\Gamma_\sigma$ but this is not the case. The width of $\sigma$ from  
fits to $|\overline S|^2\Sigma$ is a narrow 110 or 76 MeV. The  
width of $\sigma$ from fits to $|S|^2\Sigma$ is much broader at 217  
or 233 MeV. We obtain a puzzling result that the width of $\sigma$  
depends on nucleon transversity.

It is an accepted view that hadron resonance is a particle that,  
upon its production in a hadronic reaction, freely propagates and  
decays independently of other particles involved in the reaction.  
Thus its width cannot depend on the spin of recoil nucleon. However,  
so far this assumption has not been tested in experiments as all  
meson widths were determined from fits to spin-averaged mass  
distributions. This work is the first attempt to determine resonance  
parameters from different spin dependent mass distributions. It is  
therefore necessary to determine whether the apparent dependence of  
$\Gamma_\sigma$ on nucleon transversity is an artifact created by  
separate fits to $|\overline S|^2\Sigma$ and $|S|^2\Sigma$, or  
whether it is a genuine property of resonances. To this end we turn  
to simultaneous fits of $|\overline S|^2\Sigma$ and $|S|^2\Sigma$  
examining these two possibilities. Finally we note that we have  
performed separate fits to $|\overline S|^2\Sigma$ and $|S|^2\Sigma$  
obtained by $\chi^2$ minimization method (Fig.~2d). The fits are  
very similar to results for Monte Carlo method (Fig.~4a) but the  
$\chi^2/dpt$ is larger for all solutions.

\section{Simultaneous fit to $|\overline S|^2\Sigma$ and  
$|S|^2\Sigma$ with a single $\sigma$ pole}

Hadron resonance in a production process is represented as a pole  
in helicity or transversity amplitudes with mass and width  
independent of helicities or transversities of external particles.  
In this Section we examine the possibility that the same $\sigma$  
resonance contributes to both transversity amplitudes $\overline F$  
and $F$ (see (3.8)), and that the difference in the shape of  
$|\overline S|^2\Sigma$ and $|S|^2\Sigma$ is due to difference in  
coherent background. Thus we will try to show that the apparent  
dependence of $\Gamma_\sigma$ on nucleon spin is an artifact of  
separate fits to $|\overline S|^2\Sigma$ and $|S|^2\Sigma$  
distributions.

We use the parametrization (3.9) for $|\overline S|^2\Sigma$ and  
$|S|^2\Sigma$ with common $m_\sigma$ and $\Gamma_\sigma$ and use the  
programme FUMILI\cite{silin85} to perform a simultaneous fit to  
data on $|\overline S|^2\Sigma$ and $|S|^2\Sigma$. The two solutions  
for $|\overline S|^2\Sigma$ are labeled $\overline 1$ and  
$\overline 2$, while the two solutions for $|S|^2\Sigma$ are labeled  
1 and 2. Thus there are 4 simultaneous fits $(\overline 1, 1),  
(\overline 1, 2), (\overline 2, 1)$ and $(\overline 2, 2)$.  
Initially the $f_0(980)$ was fixed with a mass $m_f = 980$ MeV and  
width $\Gamma_f = 48$ MeV. An improved fit was obtained when $m_f$  
and $\Gamma_f$ were allowed to be free parameters. The best fit was  
obtained with fitted $f_0(980)$ using as initial values of  
parameters $B_i, C_i, \overline B_i, \overline C_i, i = 1,2$ the  
output values from the fit with fixed $f_0(980)$. The results are  
shown in Fig.~4b. The fitted values of $m_\sigma, \Gamma_\sigma,  
m_f$ and $\Gamma_f$ and the associated $\chi^2/dpt$ for each  
combination of solution are given in Table II.

We see in Fig.~4b that the simultaneous fits are similar to  
separate fits in Fig.~4a. Also, the cross-fits $(\overline 1 2)$ and  
$(\overline 2 1)$ are similar to fits $(\overline 1 1)$ and  
$(\overline 2 2)$, respectively. In Table II we notice that all fits  
have similar values of $m_\sigma, \Gamma_\sigma, m_f$ and  
$\Gamma_f$. The average values of mass and width of $\sigma$  
resonance are

\begin{equation}
m_\sigma = 775 \pm 17\ {\rm MeV}\ ,\ \Gamma_\sigma = 147 \pm 33\  
{\rm MeV}
\end{equation}

\noindent
This compares with $m_\sigma = 753 \pm 19$ MeV and $\Gamma_\sigma =  
108 \pm 53$ MeV obtained in\cite{svec97b} from fits only to  
$|\overline S|^2\Sigma$ without the interference with $f_0(980)$.  
The simultaneous fit of $|\overline S|^2\Sigma$ and $|S|^2\Sigma$  
and the interference with $f_0(980)$ thus require a slightly higher  
mass and a broader width of $\sigma$ state.

We now compare the $\chi^2/dpt$ for the separate and simultaneous fits:

\narrowtext
\begin{quasitable}
\begin{tabular}{lccccc}
Fit & $(\overline 1 1)$ & $(\overline 1 2)$ & $(\overline 2 1)$ &  
$(\overline 2 2)$ & Average\\
\tableline
Separate & 0.250 & 0.177 & 0.267 & 0.194 & 0.222\\
Simultaneous & 0.386 & 0.292 & 0.370 & 0.272 & 0.330
\end{tabular}
\end{quasitable}

\noindent
In this table the $\chi^2/dpt$ is the average value of $\chi^2/dpt$  
for solutions $\overline i$ and $j$, $\overline i, j = 1,2$  
calculated by the program FUMILI. We see that $\chi^2/dpt$ for  
simultaneous fits is systematically larger than for separate fits.  
The average $\chi^2/dpt$ over all combinations is 0.222 for separate  
fits and 0.330 for simultaneous fits, or 50\% larger. To reject the  
hypothesis that $\Gamma_\sigma$ depends on nucleon spin we would  
need a simultaneous fit with average $\chi^2/dpt$ over all  
combinations smaller than 0.222. Since this is not the case, we will  
explore in the next section the possibility that $\Gamma_\sigma$  
depends on nucleon spin in another simultaneous fit to $|\overline  
S|^2\Sigma$ and $|S|^2\Sigma$

At this point we comment on our use of $\chi^2$. Let $N$ be the  
number of observations, $M$ the number of free parameters and  
$\nu=N-M$ the number of degree of freedom. Then $\chi^2$ per degree  
of freedom, $\chi^2/dof = \chi^2/\nu$, is used to decide the  
goodness of the fit\cite{bevington69}. The fit is said to be good if  
$\chi^2/dof \appls 1$. In comparing various fits it is more  
appropriate to use $\chi^2$ per data point, $\chi^2/dpt = \chi^2/N$,  
calculated by the program FUMILI. $\chi^2/dpt$ reflects the  
absolute value of $\chi^2$ obtained and fits with lower $\chi^2/dpt$  
are said to be better than fits with higher $\chi^2/dpt$. Notice  
that $\chi^2/dof = (\chi^2/dpt) (N/\nu)$. We have verified that all  
our fits are good with $\chi^2/dof < 1$.

\section{Simultaneous fit to $|\overline S|^2\Sigma$ and  
$|S|^2\Sigma$ with two $\sigma$ poles}

\subsection{Two-pole parametrization}

Let us accept that the amplitude $|\overline S|^2\Sigma$ has a  
$\sigma$ pole with a narrower width and that the amplitude  
$|S|^2\Sigma$ has a $\sigma$ pole with a broader width. We can  
express the helicity amplitude in terms transversity amplitudes.  
Using the notation\cite{svec97b}

\begin{equation}
|\overline S|^2\Sigma = q |\overline F|^2 = q {1\over 2} |F_0 - iF_1|^2
\end{equation}

\[
|S|^2\Sigma = q |F|^2 = q {1\over 2} |F_0 + iF_1|^2
\]

\noindent
the helicity nonflip and flip amplitudes can be written

\begin{equation}
F_0 = {1\over{\sqrt 2}} (F + \overline F)
\end{equation}

\[
F_1 = {{-i}\over{\sqrt 2}} (F - \overline F)
\]

\noindent
The helicity amplitudes acquire two $\sigma$ poles from the  
transversity amplitudes $F$ and $\overline F$. Let us tentatively  
label $m_{\overline\sigma}$ and $\Gamma_{\overline\sigma}$ the mass  
and width of the $\overline\sigma$ pole in amplitude $|\overline  
S|^2\Sigma$, and $m_\sigma$ and $\Gamma_\sigma$ the mass and width  
of the $\sigma$ pole in the amplitude $|S|^2\Sigma$. Inverting the  
relationship (5.2) we get

\begin{equation}
F = {1\over{\sqrt 2}} (F_0 + i F_1)
\end{equation}

\[
\overline F = {1\over{\sqrt 2}} (F_0 - i F_1)
\]

\noindent
The self consistency now requires that the transversity amplitudes  
$F$ and $\overline F$ also possess the two sigma poles  
$\overline\sigma$ and $\sigma$. We could imagine that while  
$\overline\sigma$ dominates $|\overline S|^2\Sigma$, the $\sigma$  
presents a small contribution. Similarly $\sigma$ could dominate  
$|S|^2\Sigma$ with $\overline\sigma$ being a small contribution. If  
this were the case then the apparent dependence of $\sigma$ width on  
nucleon spin could be explained in terms of two sigma poles. Such  
explanation would still be in accord with the standard picture of  
hadron resonances provided that the resulting $\chi^2/dpt$ would be  
smaller than the $\chi^2/dpt$ for separate fits.

Consequently we will write the amplitudes $\overline F$ and $F$ in  
the following form:

\begin{equation}
\overline F = \overline R_{\overline\sigma} BW_{\overline\sigma} +  
\overline R_\sigma BW_\sigma + \overline R_f BW_f + \overline Q
\end{equation}

\[
F = R_\sigma BW_\sigma + R_{\overline\sigma} BW_{\overline\sigma} +  
R_f BW_f + Q
\]

\noindent
where $BW_{\overline\sigma}$, $BW_\sigma$, $BW_f$ are Breit-Wigner  
amplitudes for $\overline\sigma$, $\sigma$ and $f$ resonances and  
$\overline Q$, $Q$ are background terms. Since we expect the terms  
$BW_{\overline\sigma}$ and $BW_\sigma$ to be dominant in $|\overline  
S|^2\Sigma$ and $|S|^2\Sigma$, respectively, we can factor out the  
residue terms $\overline R_{\overline\sigma}$ and $R_\sigma$

\begin{equation}
\overline F = |\overline R_{\overline\sigma}| \{  
BW_{\overline\sigma} + \overline D BW_\sigma + \overline C BW_f +  
\overline B \} e^{i\overline\phi}
\end{equation}

\[
F = |R_\sigma| \{ BW_\sigma + D BW_{\overline\sigma} + C BW_f + B\}  
e^{i\phi}
\]

\noindent
In (5.5) $\overline D, D, \overline C, C, \overline B, B$ are  
complex numbers which we shall write in the general form

\begin{equation}
\overline D = \overline D_1 + i\overline D_2, \ldots , B=B_1 + iB_2
\end{equation}

\noindent
We will again assume $t$-averaged amplitudes (5.4). This makes the  
parameters (5.6) $t$-independent and we will assume also their mass  
independence. The parametrization for $|\overline S|^2\Sigma$ can be  
written

\begin{equation}
|\overline S|^2\Sigma = \overline Y_{old} + \overline Y_{new}
\end{equation}

\noindent
where the $\overline Y_{old}$ does not contain the terms involving  
the parameters $\overline D_1$ and $\overline D_2$ and has the same  
form as (3.9). With a notation

\begin{equation}
\overline H_\sigma = |BW_{\overline\sigma}|^2,\ H_\sigma =  
|BW_\sigma|^2,\ H_f = |BW_f|^2
\end{equation}

\noindent
we get

\[
\overline Y_{old} = q \overline N_S [ (1+ 2\overline w_\sigma  
\overline B_1 + 2\overline B_2) \overline H_\sigma + \overline B_1^2  
+ \overline B_2^2
\]

\[
(\overline C_1^2 + \overline C_2^2 ) H_f + 2 \{ (\overline w_\sigma  
\overline H_\sigma + \overline B_1 ) (w_f \overline C_1 - \overline  
C_2)
\]

\begin{equation}
+ (\overline H_\sigma + \overline B_2) (\overline C_1 + w_f  
\overline C_2)\} H_f]
\end{equation}

\begin{equation}
\overline Y_{new} = q \overline N_S [2 \{ \overline w_\sigma  
\overline H_\sigma + \overline B_1) + (\overline C_1 w_f - \overline  
C_2) H_f\} \{ \overline D_1 w_\sigma - \overline D_2\} H_\sigma +
\end{equation}

\[
+ 2 \{ (\overline H_\sigma + \overline B_2) + (\overline C_1 + w_f  
\overline C_2) H_f\} \{ \overline D_1 + \overline D_2 w_\sigma\}  
H_\sigma
\]

\[
 + \{ \overline D_1^2 + \overline D_2^2\} H_\sigma ]
\]

\noindent
In (5.9) and (5.10) the coefficients $\overline w_\sigma$,  
$w_\sigma$ and $w_f$ are defined in (3.6) for $R = \overline\sigma$,  
$\sigma$ and $f$. The normalization term $\overline N_S =  
|\overline R_{\overline\sigma}|^2$.

The parametrization for $|S|^2\Sigma$ is similar to (5.9) and  
(5.10) with some changes: the free parameters are without bars, and  
the role $\overline H_\sigma$, $\overline w_\sigma$ is interchanged  
with $H_\sigma$ and $w_\sigma$.

\subsection{Results}

We have fitted the parametrizations (5.9) and (5.10) for  
$|\overline S|^2\Sigma$ and $|S|^2\Sigma$ simultaneously to the data  
on $|\overline S|^2\Sigma$ and $|S|^2\Sigma$. Since there are two  
solutions for each distributions $|\overline S|^2\Sigma$ and  
$|S|^2\Sigma$, we obtain 4 independent fits which we label  
$(\overline 1 1), (\overline 1 2), (\overline 2 1)$ and $(\overline  
2 2)$.

We started with the expectation that the poles $\overline\sigma$ in  
$|\overline S|^2\Sigma$ and $\sigma$ in $|S|^2\Sigma$ will not  
change much, and that $\sigma$ and $\overline\sigma$ will be only  
small contributions to $|\overline S|^2\Sigma$ and $|S|^2\Sigma$,  
respectively. The actual results were different and surprising.

We run the optimization program with well over 20 initial  
conditions. In each case the fits to combinations $(\overline 2 1)$  
and $(\overline 2 2)$ converged to the same solutions for resonance  
parameters $m_{\overline\sigma}$, $\Gamma_{\overline\sigma}$,  
$m_\sigma$ and $\Gamma_\sigma$. Since these solutions are very  
similar, we call them Solution A. The fits to combination  
$(\overline 1 1)$ resulted in a different solution for the resonance  
parameters which we call Solution B. The fits to combination  
$(\overline 1 2)$ produced two solutions very close to solutions A  
and B. In each combination $(\overline i j)$ the solution for  
resonance parameters produced always the same $\chi^2/dpt$.

Even with additional runs we failed to find Solution A in the  
combination $(\overline 1 1)$ in free unconstrained fits. However  
when we run a constrained fit with $m_{\overline\sigma}$,  
$\Gamma_{\overline\sigma}$, $m_\sigma$ and $\Gamma_\sigma$ from the  
Solution A from fits to $(\overline 1 2)$, $(\overline 2 1)$ and  
$(\overline 2 2)$ combinations, we obtained excellent fits with  
$\chi^2/dpt$ nearly equal to $\chi^2/dpt$ for Solution B from  
combination $(\overline 1 1)$. The resonance parameters from the fit  
to $(\overline 1 2)$ produced the closest $\chi^2/dpt$. We have  
therefore accepted this constrained fit as the Solution A for the  
$(\overline 1 1)$ combination.

Similarly we were unable to find Solution B in unconstrained fits  
to combinations $(\overline 2 1)$ and $(\overline 2 2)$. When we run  
a constrained fit with $m_{\overline\sigma}$,  
$\Gamma_{\overline\sigma}$, $m_\sigma$ and $\Gamma_\sigma$ from the  
Solution B from $(\overline 1 1)$ combination, we obtained  
acceptable fits with somewhat larger $\chi^2/dpt$ in comparison with  
$\chi^2/dpt$ from Solution A. On the basis of $\chi^2/dpt$ alone  
the Solution A is much preferable.

The fits from the Solution A and Solution B are shown in Fig.~4c  
and Fig.~4d, respectively. The figures show two fits to each  
solution of amplitudes $|\overline S|^2\Sigma$ and $|S|^2\Sigma$.  
One fit comes from the diagonal combinations $(\overline 1 1)$ or  
$(\overline 2 2)$ (solid lines) while the other is from the  
off-diagonal combinations $(\overline 1 2)$ or $(\overline 2 1)$  
(dashed lines). In Fig.~4c and Fig.~4d we see that the two $\sigma$  
poles do not result in two peaks in the $|\overline S|^2\Sigma$  
spectrum. However, there is a hump at 670 MeV in $|S|^2\Sigma$ in  
Solution A (Fig.~4c) and a small peak at 710 MeV in $|S|^2\Sigma$ in  
Solution B (Fig.~4d).

Our initial expectation was that the $\overline\sigma$ and $\sigma$  
resonances will dominate the $|\overline S|^2\Sigma$ and  
$|S|^2\Sigma$ amplitudes, respectively. Two features of  
$\overline\sigma$ and $\sigma$ poles have emerged in all fits.  
First, $\overline\sigma$ has always a higher mass and a broader  
width than $\sigma$. Second and surprising feature is that  
$\overline\sigma$ is always the dominant contribution in all fits to  
both amplitudes $|\overline S|^2\Sigma$ and $|S|^2\Sigma$ while  
$\sigma$ presents a weaker contribution to both amplitudes. Thus our  
initial expectation that the amplitudes $|\overline S|^2\Sigma$ and  
$|S|^2\Sigma$ are dominated by a narrow $\overline\sigma$ and a  
broad $\sigma$ states is not validated. Instead, both mass spectra  
of $|\overline S|^2\Sigma$ and $|S|^2\Sigma$ can be described in  
terms of dominant $\overline\sigma$ and subdominant $\sigma$. This  
parametrization with two new $\sigma$ poles leads to $\chi^2/dpt$  
that is lower than $\chi^2/dpt$ for separate fits (see below). Thus  
the hypothesis that the width of $\sigma$ depends on nucleon spin  
must be abandoned as an artifact of separate fits.

Since $\overline\sigma$ and $\sigma$ poles are no longer associated  
with the nucleon transversities of the amplitudes $|\overline  
S|^2\Sigma$  and $|S|^2\Sigma$, we will relabel these states in a  
more conventional way. We will make the following change of notation

\[
\overline\sigma \to \sigma ,\ m_{\overline\sigma} \to m_\sigma,\  
\Gamma_{\overline\sigma} \to \Gamma_\sigma
\]

\begin{equation}
\sigma \to \sigma^\prime ,\ m_\sigma \to m_{\sigma^\prime},\  
\Gamma_\sigma \to \Gamma_{\sigma^\prime}
\end{equation}

\noindent
The results for $m_\sigma$, $\Gamma_\sigma$ (old  
$m_{\overline\sigma}$, $\Gamma_{\overline\sigma})$ and  
$m_{\sigma^\prime}$, $\Gamma_{\sigma^\prime}$ (old $m_\sigma$,  
$\Gamma_\sigma$) are shown in the Table III. The average values of  
resonance parameters over the combinations give for Solution A

\begin{equation}
m_\sigma = 786 \pm 24\ {\rm MeV},\ \Gamma_\sigma = 130 \pm 47\ {\rm MeV}
\end{equation}

\[
m_{\sigma^\prime} = 670 \pm 30\ {\rm MeV},\ \Gamma_{\sigma^\prime}  
= 59 \pm 58\ {\rm MeV}
\]

\noindent
Solution B is similar but the differences between the masses of  
$\sigma$ and $\sigma^\prime$ are smaller. The average values of  
resonance parameters over the combinations are

\begin{equation}
m_\sigma = 768 \pm 24\ {\rm MeV},\ \Gamma_\sigma = 128 \pm 53\ {\rm MeV}
\end{equation}

\[
m_{\sigma^\prime} = 711 \pm 44\ {\rm MeV},\ \Gamma_{\sigma^\prime}  
= 70 \pm 78\ {\rm MeV}
\]

The Table III also shows the $\chi^2/dpt$ for each combination  
$(\overline i j)$ which is the average of $\chi^2/dpt$ for  
$|\overline S|^2\Sigma$ and $|S|^2\Sigma$ solutions. The lowest  
$\chi^2/dpt$ is always obtained for the $(\overline 2 2)$  
combination. The largest $\chi^2/dpt$ is always obtained for the  
$(\overline 1 1)$ combination. It is worthwhile to summarize the  
$\chi^2/dpt$ of various fits

\narrowtext
\begin{quasitable}
\begin{tabular}{lccccc}
Fit & $(\overline 1 1)$ & $(\overline 1 2)$ & $(\overline 2 1)$ &  
$(\overline 2 2)$ & Average\\
\tableline
Separate & 0.250 & 0.177 & 0.267 & 0.194 & 0.222\\
Simult. $(1\sigma)$ & 0.386 & 0.292 & 0.370 & 0.272 & 0.330\\
Simult. $(2\sigma)$\\
-- Solution A & 0.274 & 0.168 & 0.172 & 0.110 & 0.181\\
-- Solution B & 0.268 & 0.168 & 0.246 & 0.180 & 0.216
\end{tabular}
\end{quasitable}

\noindent
Clearly the best overall fit is the simultaneous fit with two  
$\sigma$ poles $\sigma$ and $\sigma^\prime$, Solution A. Its  
$\chi^2/dpt$ is lower than $\chi^2/dpt$ for separate fits, and much  
lower than the $\chi^2/dpt$ for simultaneous fits with a single  
$\sigma$ pole. From this we can conclude that the hypothesis that  
the width of $\sigma$ depends on nucleon spin has been invalidated.  
Equally invalidated is the hypothesis of a single $\sigma$ pole  
contributing to both $|\overline S|^2\Sigma$ and $|S|^2\Sigma$  
amplitudes. Since the Solution B has a larger $\chi^2/dpt$ close to  
$\chi^2/dpt$ for separate fits, we propose to reject it as  
unphysical. As the result of this $\chi^2$ criterion we are left  
with two $\sigma$ states $\sigma$ and $\sigma^\prime$ from the  
Solution A. A possible physical interpretation of these states will  
be discussed in Section VI.

\subsection{Spin dependence of free parameters.}

It is of interest to examine the dependence of free parameters on  
nucleon spin. In this discussion we will restrict attention mostly  
to the Solution A.

A remarkable feature of all fits is that in all amplitudes the  
contribution of the higher mass and broader $\sigma$ dominates the  
contribution of $\sigma^\prime$. In Table IV we present the real and  
imaginary parts $D_1$ and $D_2$ of the coupling of $\sigma^\prime$  
Breit-Wigner amplitude. We see that for recoil transversity ``up''  
amplitudes $\overline 1$ and $\overline 2$, the real part $\overline  
D_1$ is negative and imaginary part $\overline D_2$ is positive.  
The opposite is true for $D_1$ and $D_2$ for the recoil nucleon  
transversity ``down'' amplitudes 1 and 2. The average absolute  
magnitude $|D|_{av}$ also depends on recoil nucleon transversity:

\narrowtext
\begin{quasitable}
\begin{tabular}{lcccc}
$|D|_{av}$ & $\overline 1$ & $\overline 2$ & $1$ & $2$\\
\tableline
Solution A & 0.08 & 0.40 & 0.62 & 0.60 \\
Solution B & 0.18 & 0.35 & 0.61 & 0.61
\end{tabular}
\end{quasitable}

\noindent
We see that the coupling of $\sigma^\prime$ is weakest in the  
amplitude $|\overline S|^2\Sigma$, Solution 1. It is strongest in  
spin down amplitude $|S|^2\Sigma$ in both solutions.

In Table V we show the real and imaginary parts $C_1$ and $C_2$ of  
the coupling of $f_0(980)$ together with the fitted values of $m_f$  
and $\Gamma_f$. We see that $\overline C_1$ and $\overline C_2$ are  
both positive in both solutions for spin up amplitude $|\overline  
S|^2\Sigma$. In contrast, $C_1$ is negative in both solutions for  
spin down amplitude $|S|^2\Sigma$. The fitted values of $m_f$ and  
$\Gamma_f$ are in agreement with expectations for $f_0(980)$.

Finally, in Table VII we present the real and imaginary parts $B_1$  
and $B_2$ of coherent background and the overall normalization  
constant. $\overline B_1$ and $\overline B_2$ are both positive for  
both solutions of spin up amplitude $|\overline S|^2\Sigma$. For  
spin down amplitude $|S|^2\Sigma$ the real part $B_1$ is negative  
while the imaginary part $B_2$ remains positive. The values of  
normalization constant $N_S$ show that $\sigma$ couples most  
strongly to Solution 1 of the spin up amplitude $|\overline  
S|^2\Sigma$.

\section{Possible constituent interpretatin of the scalar states  
$\sigma^\prime(670)$ and $\sigma(786)$.}

In the usual quark model meson resonances are $q\overline q$  
states. The mass $M$ of the $q\overline q$ states increases with its  
angular momentum $L$ as $M = M_0(2n + L)$ where $n$ is the degree  
of radial excitation. The lowest mass scalar mesons are ${}^3 P_0$  
states with masses expected to be around 1000 MeV or higher. The  
masses of $\sigma^\prime(670)$ and $\sigma(786)$ are too low to be  
$q\overline q$ states.

The fact that $\sigma^\prime(670)$ and $\sigma(786)$ are not  
observed in reactions $\gamma\gamma \to \pi^+\pi^-$ and  
$\gamma\gamma\to\pi^0\pi^0$ suggests a gluonium interpretation of  
these states. Since gluons do not couple directly to photons, we  
expect the $\sigma^\prime$ and $\sigma$ states to be absent in  
$\gamma\gamma\to\pi\pi$ reactions. This conclusion is supported by  
PLUTO and DELCO data\cite{berger84,aihara86} and more recent results  
on $\gamma\gamma\to\pi^0\pi^0$\cite{marsiske90}.

Ellis and Lanik discussed the couplings of scalar gluonium $\sigma$  
on the basis of the low-energy theorems of broken chiral symmetry  
and scale invariance, implemented using a phenomenological  
Lagrangian\cite{ellis85}. They obtained for $\sigma\to\pi^+\pi^-$  
decay the following partial width

\begin{equation}
\Gamma(\sigma\to\pi^+\pi^-) = {{(m_\sigma)^2}\over{48\pi G_0}}
\end{equation}

\noindent
where $G_0 = <0|(\alpha_s/\pi) F_{\mu\nu} F^{\mu\nu} |0>$ is the  
gluon condensate term\cite{shifman79} parametrizing the  
nonperturbative effects of QCD. The numerical values were estimated  
by the ITEP group\cite{shifman79} to be $G_0 \approx 0.012$  
(GeV)$^4$ or up to $G_0 \approx 0.030$ (GeV)$^4$ in later  
calculations\cite{bell81,bradley81}. Several recent estimates of  
$G_0$ all agree on the values around $G_0 \approx 0.020$  
(GeV)$^4$\cite{narison96,bertolini96,fabbrichesi96}.

It is very interesting to note that when we take $G_0 = 0.015$  
(GeV)$^4$, the Ellis-Lanik theorem (6.1) predicts  
$\Gamma(\sigma\to\pi^+\pi^-) = 133$ MeV for the $\sigma$ mass  
$m_\sigma = 786$ MeV and $\Gamma(\sigma^\prime \to\pi^+\pi^-)=60$  
MeV for the $\sigma^\prime$ mass $m_{\sigma^\prime} = 670$ MeV. This  
result is in perfect agreement with Eq.~(5.12) where $\Gamma_\sigma  
= 130 \pm 47$ MeV and $\Gamma_{\sigma^\prime} = 59 \pm 58$ MeV.  
However, $\Gamma_\sigma$ and $\Gamma_{\sigma^\prime}$ are the full  
widths of $\sigma(786)$ and $\sigma^\prime(670)$, so these results  
on $\Gamma(\sigma\to\pi^+\pi^-)$ and  
$\Gamma(\sigma^\prime\to\pi^+\pi^-)$ represent upper limits. When we  
use $G_0 = 0.020$ (GeV)$^4$, which is the average of the latest  
values for $G_0$\cite{narison96,bertolini96,fabbrichesi96}, we get  
for the partial width $\Gamma(\sigma\to\pi^+\pi^-) = 0.75  
\Gamma_\sigma$ and $\Gamma(\sigma^\prime \to\pi^+\pi^-) = 0.75  
\Gamma_{\sigma^\prime}$ with a very reasonable branching fraction of  
75\% for the $\pi^+\pi^-$ channel. From this agreement with  
Ellis-Lanik theorem we can conclude that $\sigma^\prime(670)$ and  
$\sigma(786)$ are both best understood as the lowest mass scalar  
gluonium states $0^{++} (gg)$.

For further interpretation of the $\sigma^\prime(670)$ and  
$\sigma(786)$ states we turn to Bjorken's classification of gluonium  
states with no spatial  
excitation\cite{bjorken79,mosher80,bjorken80}. In our discussion we  
will follow Ref.~\cite{flamm82}.

Consider gluon field strength tensor

\begin{equation}
F^\alpha_{\mu\nu} = \partial_\mu A^\alpha_\nu - \partial_\nu  
A^\alpha_\mu + g_s f_{\alpha\beta\gamma} A^\beta_\mu A^\gamma_\nu
\end{equation}

\noindent
where $A^\alpha_\mu$ are QCD gluon gauge fields. The tensor  
$F^\alpha_{\mu\nu}$ transforms like an octet under SU(3)$_c$ colour  
symmetry group. In analogy to electrodynamics we define colour  
electric fields $E^\alpha_i$ and colour magnetic fields  
$B^\alpha_i$.

\begin{equation}
E^\alpha_i = F^\alpha_{0i}\ ,\ B^\alpha_i = {1\over 2}  
\epsilon_{ijk} F^\alpha_{jk}
\end{equation}

\noindent
The colour electric fields correspond to $J^{PC} = 1^{--}$ states  
while the colour magnetic fields correspond to $J^{PC} = 1^{+-}$  
states\cite{flamm82}. Since gluons are Bose particles, their total  
wave function must be symmetric with respect to the interchange of  
all labels. We restrict our consideration to $s$-waves with no  
spatial excitations which have a symmetric space wave function. Thus  
the product of the spin and colour wave functions must be also  
symmetric. Since bound gluons are off-shell, we must also admit the  
longitudinal spin component. Then there are two distinct modes of  
lowest mass scalar gluonium $0^{++}(gg)$\cite{flamm82}: colour  
electric mode gluonium

\begin{equation}
|0^{++} EE> = \sum\limits^8_{\alpha,\beta=1} \sum\limits^3_{i,j=1}  
\delta_{\alpha\beta} \delta_{ij} |E^\alpha_i E^\beta_j>
\end{equation}

\noindent
and colour magnetic mode gluonium

\begin{equation}
|0^{++} MM> = \sum\limits^8_{\alpha,\beta = 1}  
\sum\limits^3_{i,j=1} \delta_{\alpha\beta} \delta_{ij} |B^\alpha_i  
B^\beta_j>
\end{equation}

\noindent
In (6.4) and (6.5) the indices $\alpha,\beta=1, \ldots , 8$ stand  
for colour spin, while the indices $i,j=1,\ldots ,3$ stand for the  
polarization of the field which is related to the third component of  
the spin of the corresponding off-shell gluon\cite{flamm82}.

It is expected that the magnetic mode (6.5) is heavier than the  
electric mode (6.4)\cite{bjorken79,mosher80,flamm82}. Magnetic mode  
scalar gluonium is also heavier than electric mode scalar gluonium  
in MIT bag model\cite{jaffe76}. On the basis of these considerations  
we propose to identify the lower mass $\sigma^\prime(670)$ state  
with the colour electric mode $|0^{++} EE>$, and the higher mass  
$\sigma(786)$ state with the colour magnetic mode $|0^{++} MM>$ of  
the lowest mass scalar gluonium $0^{++}(gg)$.

\section{Summary.}

We have performed a new model independent amplitude analysis of  
reaction $\pi^- p\to\pi^-\pi^+ n$ at 17.2 GeV/c extending the range  
of dipion mass from 600--880 MeV of the earlier  
analysis\cite{svec96b,svec97b} to 580--1080 MeV in the present  
analysis. In this study we focused on the resonant structure of the  
$S$-wave amplitudes $|\overline S|^2\Sigma$ and $|S|^2\Sigma$ in  
this new mass range. Our purpose was to understand the role of  
$f_0(980)$ interference and to determine better the resonance  
parameters of $\sigma(750)$ resonance from fits to both transversity  
amplitudes $|\overline S|^2\Sigma$ and $|S|^2\Sigma$. This work is  
the first attempt to determine resonance parameters of a resonance  
from fits to production amplitudes of opposite nucleon transversity.  
In the absence of previous experience, we used several fitting  
procedures to understand the resonant structure of the $S$-wave  
amplitudes.

The amplitude $|\overline S|^2\Sigma$ (recoil nucleon transversity  
``up'') shows a clear resonant peak at 750 MeV below 880 MeV, an  
enhancement above 900 MeV followed by a rapid descent to a dip at  
1010 MeV. The enhancement is more pronounced in the Solution 2. The  
enhancement and the dip are explained by the interference of  
$f_0(980)$ with coherent background and the contribution from  
$\sigma(750)$ resonance. The interference of $f_0(980)$ and  
$\sigma(750)$ does not lead to a high mass and a broad width of  
$\sigma$ as suggested in Ref.~\cite{tornquist,partphys96}.

The amplitude $|S|^2\Sigma$ (recoil nucleon transversity ``down'')  
shows a broad structure and a dip at 1010 MeV. The broad structure  
is a composite of a contribution from $\sigma(750)$ resonance and  
enhancement due to interference of $f_0(980)$ with background and  
$\sigma$. The broad structure arises as the two contributions have  
comparable magnitudes. The resonance behaviour of $|S|^2\Sigma$ was  
not recognized in the earlier analyses\cite{svec96b,svec97b} because  
the mass range was limited only to 600--880 MeV. As the result the  
amplitude $|S|^2\Sigma$ was described in\cite{svec96b,svec97b} as  
``nonresonating''.

The above description of the structure of $S$-wave amplitudes was  
first reached by using separate fits to the amplitudes $|\overline  
S|^2\Sigma$ and $|S|^2\Sigma$. However the separate fits suffer from  
the difficulty that the width of $\sigma$ is narrow 110 or 76 MeV  
in Solutions 1 and 2 of amplitude $|\overline S|^2\Sigma$ but it is  
broad at 217 and 233 MeV in Solutions 1 and 2 of amplitude  
$|S|^2\Sigma$.

Next we tried to ascertain whether this apparent dependence of  
$\sigma$ width $\Gamma_\sigma$ on nucleon spin is a real physical  
effect or an artifact of separate fits to $|\overline S|^2\Sigma$  
and $|S|^2\Sigma$. To this end we have first performed a  
simultaneous fit to $|\overline S|^2\Sigma$ and $|S|^2\Sigma$ with a  
common $\sigma$ pole. The differences in the shape of mass  
distributions $|\overline S|^2\Sigma$ and $|S|^2\Sigma$ were  
expected to arise purely from the coherent background. The result of  
the best fit gives

\begin{equation}
m_\sigma = 775 \pm 17\ {\rm MeV}\ ,\ \Gamma_\sigma = 147\pm 33\ {\rm MeV}
\end{equation}

\noindent
We note that the previous determinization\cite{svec97b} of  
$m_\sigma$ and $\Gamma_\sigma$ for $|\overline S|^2\Sigma$ alone and  
without $f_0(980)$ interference gave somewhat smaller mass and a  
narrower width.

\begin{equation}
m_\sigma = 753 \pm 19\ {\rm MeV}\ ,\ \Gamma_\sigma = 108\pm 53\ {\rm MeV}
\end{equation}

The values of $\chi^2/dpt$ in the best simultaneous fit show a good  
fit but nevertheless they are higher than the $\chi^2/dpt$ from the  
best separate fits. Thus the possibility of the dependence of  
$\Gamma_\sigma$ on nucleon spin has not been eliminated by these  
fits. Consequently we turned to examine the hypothesis that the  
width $\Gamma_\sigma$ does depend on nucleon spin in a new set of  
simultaneous fits to $|\overline S|^2\Sigma$ and $|S|^2\Sigma$. If  
we denote $\overline\sigma$ the narrow $\sigma$ contributing to  
$|\overline S|^2\Sigma$ and $\sigma$ the broader $\sigma$  
contributing to $|S|^2\Sigma$, the self-consistency requires that  
some $\sigma$ contributes to $|\overline S|^2\Sigma$ and some  
$\overline\sigma$ contributes to $|S|^2\Sigma$. So instead of one  
$\sigma$ pole with a width dependent on nucleon spin, we were now  
looking for two $\sigma$ poles of similar masses with different  
widths and different relative contributions to the amplitudes  
$|\overline S|^2\Sigma$ and $|S|^2\Sigma$ to explain the results of  
separate fits.

The results of the simultaneous fits with two $\sigma$ poles turned  
out differently. We did find two $\sigma$ poles with different  
widths, but they also have considerably different masses and did not  
distinguish themselves by contributing preferentially to one or the  
other $S$-wave amplitude. We called $\sigma$ the $\sigma$ pole with  
higher mass and broader width, and $\sigma^\prime$ the $\sigma$  
pole with lower mass and narrower width. We found that $\sigma$ is  
the dominating contribution in both solutions in both amplitudes  
$|\overline S|^2\Sigma$ and $|S|^2\Sigma$. The contribution of  
$\sigma^\prime$ is weaker in all amplitudes, but it does depend on  
the nucleon spin. It is weak in the spin ``up'' amplitudes  
$|\overline S|^2\Sigma$, and much stronger in the spin ``down''  
amplitudes $|S|^2\Sigma$.

We found two solutions for the two poles $\sigma$ and  
$\sigma^\prime$. The Solution A is much prefered on the basis of  
very low $\chi^2/dpt$. The averaged resonance parameters are

\begin{equation}
m_\sigma = 786\pm 24\ {\rm MeV}\ ,\ \Gamma_\sigma = 130\pm 47\ {\rm MeV}
\end{equation}

\[
m_{\sigma^\prime} = 670\pm 30\ {\rm MeV}\ ,\ \Gamma_{\sigma^\prime}  
= 59\pm 58\ {\rm MeV}
\]

\noindent
The Solution A for $\sigma$ and $\sigma^\prime$ has $\chi^2/dpt$  
lower than $\chi^2/dpt$ for separate fits or simultaneous fits with  
single $\sigma$ pole. On the basis of this $\chi^2$ criterion we  
recommend that the Solution A for $\sigma$ and $\sigma^\prime$ poles  
be accepted as the actual physical explanation of the resonant  
structure of amplitudes $|\overline S|^2\Sigma$ and $|S|^2\Sigma$.

In Section VI we showed that both $\sigma$ and $\sigma^\prime$  
poles meet several criteria for gluonium candidates: their masses  
are too low to be a $q\overline q$ scalar states, they are not  
observed in $\gamma\gamma\to \pi\pi$ reactions, and they both  
satisfy Ellis-Lanik theorem (6.1) with the same value of gluon  
codensate $G_0$. On the basis of Bjorken's classification of  
gluonium states\cite{bjorken79,mosher80,bjorken80,flamm82} we  
proposed to identify $\sigma^\prime(670)$ and $\sigma(786)$ with the  
colour electric and colour magnetic modes of lowest mass scalar  
gluonium $0^{++} (gg)$.

The CERN measurements of pion production on polarized targets have  
opened a whole new approach to experimental hadron spectroscopy by  
making possible the study of resonance production on the level of  
spin amplitudes. Our results emphasize the need for a dedicated and  
systematic study of various production processes on the level of  
spin amplitudes measured in experiments with polarized targets.

\acknowledgements
I wish to thank N.A.~T\"ornquist for suggesting that I extend my  
earlier studies and examine the interference with $f_0(980)$. I  
thank K.~Rybicki for providing me with the numerical data of  
Ref.~\cite{chabaud83,kamins97}, which enabled this study at higher  
masses. I wish to thank him also for his interest and stimulating  
e-mail correspondence. I am also grateful to J.~Bystricky who  
explained to me how to use the program FUMILI for multifunction  
optimization needed for simultaneous fits. This work was supported  
by Fonds pour la Formation de Chercheurs et l'Aide \`a la Recherche  
(FCAR), Minist\`ere de l'Education du Qu\'ebec, Canada.

\begin{figure}
\caption{Mass dependence of physical solutions for moduli squared  
of $S$-wave and $P$-wave nucleon transversity amplitudes and cosines  
of their relative phases in reaction $\pi^- p \to \pi^- \pi^+ n$ at  
17.2 GeV/c and momentum transfers $-t=0.995-0.2$ (GeV/c)$^2$. The  
results are in the $t$-channel dipion helicity frame.}\label{fig1}
\end{figure}

\begin{figure}
\caption{Mass dependence of unnormalized amplitudes $|\overline  
S|^2\Sigma$ and $|S|^2\Sigma$ measured in $\pi^- p_\uparrow \to  
\pi^- \pi^+ n$ at 17.2 GeV/c and $-t = 0.005 - 0.20$ (GeV/c)$^2$.  
Figures (a) and (b) show the results using the Monte Carlo method  
for amplitude analysis of Data Set 1 (Ref.~[17]) and Data Set 3  
(this paper). Figures (c) and (d) show the results using the  
$\chi^2$ minimization method for amplitude analysis of Data Set 1  
(Ref.~[5,18]) and Data Set 3 (Ref.~[10]), respectively.}\label{fig2}
\end{figure}

\begin{figure}
\caption{Four solutions for the $S$-wave partial-wave intensity  
$I_S$ in the reaction $\pi^- p \to \pi^-\pi^+ n$ at 17.2 GeV/c and  
$-t=0.005-0.20$ (GeV/c)$^2$.}\label{fig3}
\end{figure}

\begin{figure}
\caption{The fits to amplitudes $|\overline S|^2\Sigma$ and  
$|S|^2\Sigma$. (a) The separate fits with separate single $\sigma$  
pole using the parametrization (3.9). (b) The simultaneous fit with  
a common single $\sigma$ pole using the parametrization (3.9).  
Figures (c) and (d) show Solution A and B, respectively, of the  
simultaneous fit with two common $\sigma$ poles using the  
parametrization (5.9) and (5.10). The Solution A in (c) gives the  
best $\chi^2/dpt$.}\label{fig4}
\end{figure}

\begin{table}
\caption{Mass and width of $\sigma(750)$ and $f_0(980)$ from  
separate fits to solutions $\overline 1$ and $\overline 2$ of  
amplitude $|\overline S|^2\Sigma$, and to solutions 1 and 2 of  
amplitude $|S|^2\Sigma$.}\label{table1}
\begin{tabular}{lccccc}
{Solution} &{$m_\sigma$} &{$\Gamma_\sigma$} &{$m_f$} &{$\Gamma_f$}  
&{$\chi^2$/dpt}\\
 &{(MeV)} &{(MeV)} & {(MeV)} & {(MeV)}\\
\tableline
$\overline 1$ & 773 $\pm$ 12 & 110 $\pm$ 25 & 956 $\pm$ 17 & 81  
$\pm$ 40 &\dec 0.206 \\
$\overline 2$ & 775 $\pm$ 16 & 76 $\pm$ 37 & 973 $\pm$ 17 & 57  
$\pm$ 25 &\dec 0.240 \\
1 & 788 $\pm$ 33 & 217 $\pm$ 53 & 982 $\pm$ 10 & 42 $\pm$ 39 &\dec  
0.294 \\
2 & 807 $\pm$ 45 & 233 $\pm$ 109 & 988 $\pm$ 10 & 16 $\pm$ 25   
&\dec 0.148 \\
\end{tabular}
\end{table}

\begin{table}
\caption{Mass and width of $\sigma(750)$ and $f_0(980)$ from  
simultaneous fits to $|\overline S|^2\Sigma$ and $|S|^2\Sigma$ with  
a common single $\sigma$ pole. The combination of solutions of  
$|\overline S|^2\Sigma$ and $|S|^2\Sigma$ is given in the  
parentheses on the left with the notation as in Table I. The  
$\chi^2/dpt$ shown is the average of $\chi^2/dpt$ for $|\overline  
S|^2\Sigma$ and $|S|^2\Sigma$ solutions.}\label{table2}
\begin{tabular}{cccccc}
{Fit} &{$m_\sigma$} &{$\Gamma_\sigma$} &{$m_f$} &{$\Gamma_f$}  
&{$\chi^2$/dpt}\\
 &{(MeV)} &{(MeV)} & {(MeV)} & {(MeV)}\\
\tableline
$(\overline 1 1)$ & 771 $\pm$ 13 & 161 $\pm$ 22 & 980 $\pm$ 10 & 49  
$\pm$ 21 &\dec 0.386 \\
$(\overline 1 2)$ & 771 $\pm$ 14 & 145 $\pm$ 25 & 970 $\pm$ 19 & 60  
$\pm$ 28 &\dec 0.292 \\
$(\overline 2 1)$ & 780 $\pm$ 21 & 156 $\pm$ 43 & 980 $\pm$ 9 & 47  
$\pm$ 25 &\dec 0.370 \\
$(\overline 2 2)$ & 780 $\pm$ 18 & 125 $\pm$ 40 & 975 $\pm$ 45 & 20  
$\pm$ 20  &\dec 0.272 \\
\end{tabular}
\end{table}

\begin{table}
\caption{Mass and width of two $\sigma$ poles from simultaneous  
fits to $|\overline S|^2\Sigma$ and $|S|^2\Sigma$ using  
parametrization (5.9), (5.10) for $|\overline S|^2\Sigma$ and  
corresponding parametrization for $|S|^2\Sigma$. Values without  
errors were obtained in a constrained fit. The $\chi^2/dpt$ shown is  
the average of $\chi^2/dpt$ for $|\overline S|^2\Sigma$ and  
$|S|^2\Sigma$ solutions.}\label{table3}
\begin{tabular}{cccccc}
{Amplitudes} &{$m_\sigma$} &{$\Gamma_\sigma$}  
&{$m_{\sigma^\prime}$} &{$\Gamma_{\sigma^\prime}$} &{$\chi^2$/dpt}\\
 &{(MeV)} &{(MeV)} & {(MeV)} & {(MeV)}\\
\tableline
\multicolumn{2}{l}{Solution A}\\
$(\overline 1 1)$ & 775 & 118 & 676 & 73 &\dec 0.274 \\
$(\overline 1 2)$ & 775 $\pm$ 19 & 118 $\pm$ 34 & 676 $\pm$ 44 & 73  
$\pm$ 71 &\dec 0.168 \\
$(\overline 2 1)$& 779 $\pm$ 27 & 167 $\pm$ 63 & 666 $\pm$ 12 & 36  
$\pm$ 38 &\dec 0.172 \\
$(\overline 2 2)$ & 795 $\pm$ 29 & 120 $\pm$ 57 & 665 $\pm$ 18 & 55  
$\pm$ 51  &\dec 0.110 \\
\multicolumn{2}{l}{Solution B}\\
$(\overline 1 1)$ & 761 $\pm$ 27 & 140 $\pm$ 69 & 717 $\pm$ 43 & 72  
$\pm$ 79 &\dec 0.268 \\
$(\overline 1 2)$ & 770 $\pm$ 23 & 123 $\pm$ 48 & 709 $\pm$ 44 & 69  
$\pm$ 77 &\dec 0.168 \\
$(\overline 2 1)$& 770 & 123 & 709 & 69 &\dec 0.246 \\
$(\overline 2 2)$ & 770 & 123 & 709 & 69  &\dec 0.180 \\
\end{tabular}
\end{table}

\begin{table}
\caption{Spin dependence of the $\sigma^\prime$ coupling constants  
$D_1$ and $D_2$ defined in (5.5). $\overline D_1$ and $\overline  
D_2$ are understood for amplitudes $\overline 1$ and $\overline  
2$.}\label{table4}
\begin{tabular}{ccccc}
{Amplitude} &${m_\sigma, \Gamma_\sigma}$ (MeV)  
&{$m_{\sigma^\prime},\Gamma_{\sigma^\prime}$ (MeV)} &{$D_1$}  
&{$D_2$}\\
\tableline
$\overline 1(\overline 1 1)$ & (775, 118) & (676, 73) & $-0.06 \pm  
0.10$ & $0.05 \pm 0.10$ \\
$\overline 1(\overline 1 2)$ & (775, 118) & (676, 73) & $-0.06 \pm  
0.20$ & $0.06 \pm 0.15$ \\
$\overline 2(\overline 2 1)$& (799, 167) & (666, 36) & $-0.36 \pm  
0.36$ & $0.20 \pm 0.25$ \\
\bigskip
$\overline 2(\overline 2 2)$ & (795, 120) & (665, 55) & $-0.31 \pm  
0.36$ & $0.25 \pm 0.29$ \\
$1(\overline 1 1)$ & (775, 118) & (676, 73) & $0.80 \pm 0.60$ &  
$-0.38 \pm 0.53$ \\
$1(\overline 2 1)$ & (799, 167) & (666, 36) & $-0.06 \pm 0.33$ &  
$-0.34 \pm 0.36$ \\
$2(\overline 1 2)$& (775, 118) & (676, 73) & $0.65 \pm 0.88$ &  
$-0.27 \pm 1.37$ \\
$2(\overline 2 2)$ & (795, 120) & (665, 55) & $0.10 \pm 0.52$ &  
$-0.47 \pm 0.56$ \\
\end{tabular}
\end{table}

\begin{table}
\caption{Spin dependence of the $f_0(980)$ coupling constants $C_1$  
and $C_2$ defined in (5.5) and the fitted values for $m_f$ and  
$\Gamma_f$. $\overline C_1$ and $\overline C_2$ are understood for  
amplitudes $\overline 1$ and $\overline 2$. The ordering of masses  
is as in Table IV.}\label{table5}
\begin{tabular}{ccccc}
{Amplitude} & $C_1$ & $C_2$ & $m_f$ (MeV) & $\Gamma_f$ (MeV)\\
\tableline
$\overline 1(\overline 1 1)$ & $0.15 \pm 0.28$ & $0.65 \pm 0.20$ &  
$968 \pm 16$ & $95 \pm 29$ \\
$\overline 1(\overline 1 2)$ & $0.30 \pm 0.55$ & $0.60 \pm 0.26$ &  
$958 \pm 18$ & $91 \pm 40$ \\
$\overline 2(\overline 2 1)$ & $0.92 \pm 1.04$ & $1.34 \pm 1.44$ &  
$981 \pm 7$ & $35 \pm 24$ \\
\bigskip
$\overline 2(\overline 2 2)$ & $1.38 \pm 1.05$ & $0.89 \pm 1.41$ &  
$973 \pm 16$ & $58 \pm 26$ \\
$1(\overline 1 1)$ & $-1.37 \pm 0.71$ & $-0.21 \pm 0.86$ & $968 \pm  
16$ & $95 \pm 29$ \\
$1(\overline 2 1)$ & $-0.83 \pm 0.45$ & $0.35 \pm 0.56$ & $981 \pm  
7$ & $34 \pm 24$ \\
$2(\overline 1 2)$ & $-1.09 \pm 1.29$ & $0.33 \pm 1.30$ & $958 \pm  
19$ & $91 \pm 41$ \\
$2(\overline 2 2)$ & $-0.66 \pm 0.85$ & $0.56 \pm 0.71$ & $974 \pm  
15$ & $58 \pm 26$ \\
\end{tabular}
\end{table}

\begin{table}
\caption{Spin dependence of coherent background $B_1$ and $B_2$ and  
the normalization constant $N_S$ defined in (5.5), (5.9) and  
(5.10). $\overline B_1$, $\overline B_2$ and $\overline N_S$ are  
understood for amplitudes $\overline 1$ and $\overline 2$. The  
ordering of masses as in Table IV.}\label{table6}
\begin{tabular}{cccc}
{Amplitude} & $B_1$ & $B_2$ & $N_S$\\
\tableline
$\overline 1(\overline 1 1)$ & $0.82 \pm 0.19$ & $0.27 \pm 0.13$ &  
$2.81 \pm 0.74$ \\
$\overline 1(\overline 1 2)$ & $0.81 \pm 0.29$ & $0.22 \pm 0.35$ &  
$2.91 \pm 0.91$ \\
$\overline 2(\overline 2 1)$ & $1.59 \pm 0.64$ & $0.36 \pm 0.75$ &  
$1.42 \pm 0.82$ \\
\bigskip
$\overline 2(\overline 2 2)$ & $1.82 \pm 0.73$ & $0.15 \pm 1.26$ &  
$1.26 \pm 0.82$ \\
$1(\overline 1 1)$ & $-0.92 \pm 0.60$ & $0.44 \pm 0.42$ & $0.76 \pm  
0.61$ \\
$1(\overline 2 1)$ & $-0.12 \pm 0.25$ & $0.35 \pm 0.44$ & $1.92 \pm  
1.56$ \\
$2(\overline 1 2)$& $-1.10 \pm 1.42$ & $0.61 \pm 0.98$ & $0.80 \pm  
1.26$ \\
$2(\overline 2 2)$ & $-0.47 \pm 0.74$ & $0.89 \pm 0.65$ & $1.15 \pm  
1.06$ \\
\end{tabular}
\end{table}

\end{document}